\documentclass[conference]{IEEEtran}
\IEEEoverridecommandlockouts
\usepackage{cite}
\usepackage{amsmath,amssymb,amsfonts}
\usepackage{algorithmic}
\usepackage{graphicx}
\usepackage{textcomp}
\usepackage{xcolor}
\def\BibTeX{{\rm B\kern-.05em{\sc i\kern-.025em b}\kern-.08em
    T\kern-.1667em\lower.7ex\hbox{E}\kern-.125emX}}
\begin{document}

\title{SpeedyChain: A framework for decoupling data from blockchain for smart cities
}

\author{\IEEEauthorblockN{1\textsuperscript{st} Regio A. Michelin}
\thanks{The first and second authors have the same contribution for the present research.}
\IEEEauthorblockA{
\textit{IFRS and PUCRS}\\
Porto Alegre, Brazil \\
regio.michelin@acad.pucrs.br}
\and
\IEEEauthorblockN{2\textsuperscript{nd} Ali Dorri}
\IEEEauthorblockA{
\textit{CSE, USNW, DATA61 and CSIRO}\\
Sydney, Australia \\
ali.dorri@unsw.edu.au}
\and
\IEEEauthorblockN{3\textsuperscript{rd} Roben C. Lunardi}
\IEEEauthorblockA{
\textit{IFRS and PUCRS}\\
Porto Alegre, Brazil \\
roben.lunardi@acad.pucrs.br}
\and
\IEEEauthorblockN{4\textsuperscript{th} Marco Steger}
\IEEEauthorblockA{
\textit{Virtual vehicle research center}\\
Austria \\
marco.steger@v2c2.at}
\and
\IEEEauthorblockN{5\textsuperscript{th} Salil S. Kanhere}
\IEEEauthorblockA{
\textit{CSE and UNSW}\\
Sydney, Australia \\
salil.kanhere@unsw.edu.au}
\and
\IEEEauthorblockN{6\textsuperscript{th} Raja Jurdak}
\IEEEauthorblockA{
\textit{DATA61 and CSIRO}\\
Brisbane, Australia \\
raja.jurdak@data61.csiro.au}
\and
\IEEEauthorblockN{7\textsuperscript{th} Avelino F. Zorzo}
\IEEEauthorblockA{
\textit{PUCRS}\\
Porto Alegre, Brazil \\
avelino.zorzo@pucrs.br}
}

\maketitle

\begin{abstract}
There is increased interest in smart vehicles acting as both data consumers and producers in smart cities. Vehicles can use smart city data for decision-making, such as dynamic routing based on traffic conditions. Moreover, the multitude of embedded sensors in vehicles can collectively produce a rich data set of the urban landscape that can be used to provide a range of services.  Key to the success of this vision is a scalable and private architecture for trusted data sharing. This paper proposes a framework called SpeedyChain, that leverages blockchain technology to allow smart vehicles to share their data while maintaining privacy, integrity, resilience and non-repudiation in a decentralized, and tamper-resistant manner. Differently from traditional blockchain usage (\textit{e.g.}, Bitcoin and Ethereum), the proposed framework uses a blockchain design that decouples the data stored in the transactions from the block header, thus allowing for fast addition of data to the blocks. Furthermore, an expiration time for each block to avoid large sized blocks is proposed. This paper also presents an evaluation of the proposed framework in a network emulator to demonstrate its benefits.
\end{abstract}

\begin{IEEEkeywords}
Blockchain, Data Decouple, Appendable Block, Smart Vehicles, Smart Cities
\end{IEEEkeywords}

\section{Introduction}\label{sec:introduction}
A smart city incorporates information and communication technology to enhance the quality of life of its citizens and improve the efficiency of urban services such as utilities, energy, and transportation. Intelligent Transportation Systems (ITS) are an integral part of smart cities. ITS use sensing, communication, analytics, and control to improve the safety and efficiency of city-wide transportation systems. Future connected vehicles interacting will be equipped with a substantial number of sensors (such as Global Positioning System (GPS), dashboard cameras, Light Detection and Ranging (LIDAR), etc.) that will produce  large volumes of data. Research conducted by Intel predicts that future vehicles will produce 4,000 GB of data every day \cite{intel}. The data produced by smart vehicles will be used by smart urban infrastructures to offer services, \textit{e.g.}, finding available parking spots~\cite{Kubler:2016} or helping in a traffic incident~\cite{Yamamoto:2016}. Thus, it is crucial to ensure the validity of the data, \textit{i.e.}, to ensure that the data are produced by a trusted vehicle. 

The significant volume of data produced by the vehicles, as well as the requirement to trust the data producer raises the demand for a low-latency trusted data exchange platform. In such a framework, participants in a smart city, including vehicles, roadside infrastructure units (RSIs), and service providers (SPs), must be able to validate the veracity of the information they receive, \textit{i.e.}, ensure that the data are produced by a legitimate entity and that these data have not been altered. Furthermore, the validation must be performed with low latency given the high volume of data and the real-time nature of some services (\textit{e.g.}, traffic updates or accident reports). 

The exchanged data between vehicles may contain privacy-sensitive information about the vehicle owner, \textit{e.g.}, the location of the vehicle. If the identity of the vehicle owner is revealed to third parties such as SPs, then this could be construed as a serious breach of personal privacy. Hence, any smart city platform must guarantee that privacy is provided as a fundamental requirement. 

Existing methods to achieve data integrity and veracity~\cite{Jin2014},~\cite{Zanella2014} suffer from the following issues: \par 
\begin{itemize}
    \item \textbf{Centralization:} most of the existing methods rely on centralized brokers, such as the vehicle manufacturer, and are unlikely to scale and accommodate millions of connected vehicles, each of which generating a large amount of sensing data ~\cite{Zanella2014}.
    \item \textbf{Lack of Privacy: } the existing data exchange platforms require a vehicle to authenticate itself to ensure trust. This compromises the vehicle owner privacy as the exchanged data is now connected to their identity~\cite{Bloom2017},~\cite{Mulligan2013}. 
    \item \textbf{Data Vulnerability: } data collected by connected vehicles is typically stored in traditional data stores (e.g., centralized Database Management Systems)~\cite{Zanella2014}, which have been shown to be vulnerable to breaches~\cite{Algarni2016}. Tampering with these data could lead to serious harm, \textit{e.g}, causing accidents if the falsified data are used to control traffic lights.
\end{itemize}

Blockchain~\cite{dorri2017towards} is a promising new technology that has the potential to address the aforementioned challenges. It supports decentralization, security, and privacy and is being widely used in diverse disciplines including finance, law, Internet of Things (IoT) and energy~\cite{Tschorsch2016}.  Blockchain maintains a distributed ledger of blocks that are shared across all participating nodes. Each transaction, \textit{i.e.}, data exchange between nodes, is verified by all, or some, participating nodes, thus eliminating the need for central authorities. Transactions are grouped in blocks and then appended to the blockchain. Appending a new block to the blockchain, known as mining, entails solving a puzzle or even voting in order to achieve a consensus among the nodes of the Peer-to-Peer (P2P) network that distributively manages the blockchain. This puzzle underpins a trustless consensus algorithm among untrusted nodes. 
Sensible data transferred between nodes can be encrypted, which protects against eavesdropping. In order to ensure the user privacy, the Bitcoin cryptocurrency \cite{Nakamoto2008}, for example, is based on a nonexistence of public association between Public Key (PK) and its owner (person).

\begin{figure}[h!]
 \begin{center}
\includegraphics[width=0.48\textwidth]{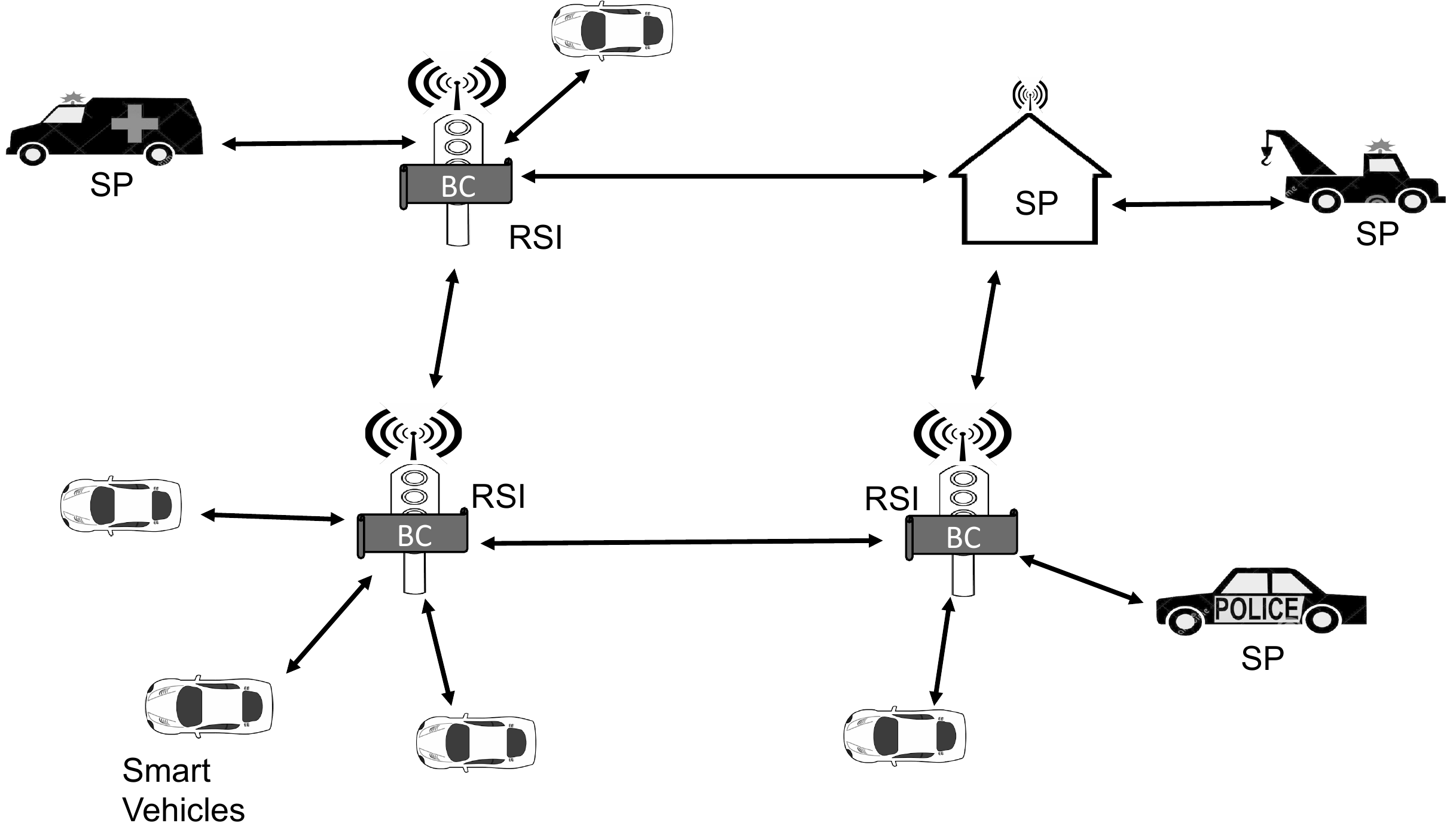}
 \end{center}
\caption{Smart city scenario.}
\label{fig:smartcity}
\end{figure}

This paper proposes a blockchain-based data exchange framework for smart cities called SpeedyChain. SpeedyChain includes smart city infrastructure (\textit{e.g.}, traffic lights), smart vehicles, RSIs, and SPs as shown in Figure~\ref{fig:smartcity}. 
The blockchain management, \textit{i.e.}, verification of transactions and blocks, and mining new blocks, are performed by participants that have both more available resources and a closer commitment to the smart city, \textit{e.g.}, RSIs and SPs.

The most prevalent blockchain implementation, Bitcoin, demands  high storage capacity as well as processing power to run the consensus algorithm called Proof-of-Work (PoW)~\cite{Nakamoto2008}. This algorithm rewards the nodes based on their processing power. Other blockchain solutions focusing on IoT devices such as IOTA~\cite{iota} rely on simplified versions of the PoW algorithm. Additionally, these traditional blockchains have a high latency when including a new transaction, which is not well suited for the smart city scenario as outlined earlier. In a previous work, Lunardi \textit{et al.}~\cite{lunardi2018} proposed a customized blockchain that relies upon a single block per device identified by its public key and storing signed transactions to address these limitations. This blockchain allows appending data to an existing block through the hash of the previous information and signing the new created information. Moreover, since a new transaction can be instantaneously added to a block, it does not suffer from the latency issues associated with the formation of a block by collating pending transactions. However, it does not support privacy and mobility aspects that are desired for smart cities.
 
By using SpeedyChain, the participants in the blockchain can establish communication with non-repudiation, data integrity and resilience, which not only allows a vehicle to be authenticated by RSIs but also allows RSIs to be trusted by vehicles. Once the trust is established, the vehicle will have a block created on the blockchain and is allowed to append transactions to his block. To ensure user privacy, our framework proposal relies on the anonymity principle that is used in Bitcoin, thus vehicles must change their keys in defined time intervals known as key update interval (KUI). Within a KUI, vehicles must replace their current key pair by a new one and update the blockchain accordingly.

The main contribution of this paper is SpeedyChain, a permissioned blockchain-based framework for ensuring resilient, decentralized and immutable management of smart city data. SpeedyChain also ensures reliable Vehicle-to-Infrastructure communication and maintains vehicle privacy by employing periodically changeable keys. Therefore, important changes were incorporated in existing blockchain based-solutions for a similar IoT scenario, especially related to the definition of expiration time of a block (that can be used to avoid large sized blocks), the inclusion of the access level for devices (that can help to manage permissions of each vehicle) and the adoption of key update algorithm (that can reduce the traceability of the vehicle data). Also, we undertake extensive experiments to evaluate the performance of SpeedyChain in an emulation environment using Common Open Research Emulator (CORE)~\cite{CORE2008}. For example, we varied the number of created transactions between 10 and 1,000 and measured the time required to validate and append a new transaction in the blockchain. In our experimental results, we observed a linear growth in time and were able to show the low latency introduced by our framework (\textit{i.e.}, under 2 milliseconds).
Consequently, the results demonstrate that through SpeedyChain usage, it is possible to exchange information in smart cities ensuring resilience, data integrity and tamper-resistance.

The rest of this paper is organized as follows. Section~\ref{sec:relatedWork} presents an overview of research related to smart cities and blockchain applications. Section~\ref{sec:background}  presents a synopsis of the underlying blockchain technology that we have used and the changes incorporated to make it applicable to the smart city scenario. Section~\ref{sec:architecture} provides the details of our proposed blockchain-based data exchange framework. A brief discussion of security and extensive evaluations are presented in Section~\ref{sec:evaluations}. Finally, we draw some conclusions in Section~\ref{sec:conclusion}.

\section{Related Work}\label{sec:relatedWork}

In most smart city scenarios, Intelligent Transportation Systems (ITS) are mostly used to provide connected vehicles with information about the current traffic situation as well as about road and weather conditions. Furthermore, ITS enable/support beneficial functions such as path planning or mechanisms to warn road users about traffic jams, approaching emergency vehicles, and dangerous roadworks. 
To do so, vehicles and roadside infrastructure units (RSIs), such as traffic lights, need to exchange data using Vehicle-to-Infrastructure (V2I) communication. One key issue in such ITS applications is the privacy of the involved users \cite{priv-ITS}. This privacy is defined as the capability of keeping users data anonymous such that they cannot be linked to their real identity.

To tackle this issue, researchers have proposed solutions \cite{priv-ITS, priv-ITS1, priv-ITS2, priv-ITS3} for classical ITS functions, which mitigate most of the known privacy issues. However, the discussed solutions do not tackle the challenges of future smart cities, like supporting a decentralized trust model where multiple entities (SPs, RSIs, and vehicles) act together to share information.

In order to identify the privacy issues related to smart vehicles, Bloom \textit{et al.}~\cite{Bloom2017} focus on people's perception about the data collected by self-driving cars. They conducted a study with 302 participants and showed that people are not aware of the extensive amount of data collected by these cars including GPS data, images, speed and how these data are shared and used. Once the participants were made aware of this fact, they  expressed concerns about how the data could be manipulated or who could have access to that data. Their findings suggested that the associated privacy concerns could have a negative impact on the acceptance of autonomous driving vehicles.

Privacy in smart cities and vehicular networks is an issue discussed in several works \cite{Papadimitratos2008, Lu2012, Hubaux2004, Hoh2012}. Papadimitratos~\cite{Papadimitratos2008}, for example, proposed an architecture that relies on a certification authority (CA) that is responsible for the identity management in its own region (like city, district, county, etc.). Thus, each region or city has its own CA, and region-to-region cross-certification is used to allow a vehicle to move from one CA to another. The centralization of the CA is a bottleneck and a single point of failure, which motivates the need for a decentralized solution. To address the privacy issue, the authors proposed the creation of a pool of pseudonyms that are assigned to vehicles when they move between regions.

\begin{table*}[!ht]
\centering
\caption{Related work aspects comparison}
\label{tab:related}
\begin{tabular}{l|c|c|c|c|c|c|}
\cline{2-7}
 & \textbf{IOTA~\cite{iota}} & \textbf{Dorri \textit{et al.}~\cite{dorri2017towards,Dorri2017}} & \textbf{Lunardi \textit{et al.}~\cite{lunardi2018}} & \textbf{Sharma \textit{et al.}~\cite{Sharma2017}} & \textbf{Li \textit{et al.}~\cite{Li2018}} & \textbf{SpeedyChain} \\ \hline
\multicolumn{1}{|l|}{\textbf{\begin{tabular}[c]{@{}l@{}}Time to add \\ transaction\end{tabular}}} & Minutes & ms to seconds & \textgreater 40 ms & N/A & \textgreater 40 ms & \textless 2 ms \\ \hline
\multicolumn{1}{|l|}{\textbf{Architecture}} & P2P & Overlays & Gateways & \begin{tabular}[c]{@{}c@{}}Vehicles and \\ Miner Nodes\end{tabular} & RSU/OBU & RSI - Vehicles \\ \hline
\multicolumn{1}{|l|}{\textbf{Hardware}} & \begin{tabular}[c]{@{}c@{}}Node: PC; Wallet: \\ Own/Rasp.Pi\end{tabular} & N/A & \begin{tabular}[c]{@{}c@{}}Arduino, Orange Pi \\ and Raspberry Pi\end{tabular} & N/A & Simulated & Emulated \\ \hline
\multicolumn{1}{|l|}{\textbf{Block}} & Immutable & Immutable & Decoupled & Immutable & Immutable & Decoupled \\ \hline
\multicolumn{1}{|l|}{\textbf{Main Usage}} & Payment M2M & \begin{tabular}[c]{@{}c@{}}Smart Homes / \\ Smart Cities\end{tabular} & \begin{tabular}[c]{@{}c@{}}Smart Homes / \\ Smart Offices\end{tabular} & Smart Cities & Smart Cities & Smart Cities \\ \hline
\multicolumn{1}{|l|}{\textbf{\begin{tabular}[c]{@{}l@{}}Key \\ Management\end{tabular}}} & \begin{tabular}[c]{@{}c@{}}One key pair\\  per device\end{tabular} & \begin{tabular}[c]{@{}c@{}}One key pair \\ per device\end{tabular} & \begin{tabular}[c]{@{}c@{}}One key pair \\ per device\end{tabular} & \begin{tabular}[c]{@{}c@{}}One key pair\\  per node\end{tabular} & \begin{tabular}[c]{@{}c@{}}One key pair\\ per node\end{tabular} & \begin{tabular}[c]{@{}c@{}}Expiration date \\ for each PubKey. \\ Only one \\ PubKey active.\end{tabular} \\ \hline
\end{tabular}
\end{table*}

Blockchain technology provides a decentralized and resilient solution through a Peer-to-Peer (P2P) network, and ensures data integrity by employing a hash of the data stored in the blockchain. Li \textit{et al.}~\cite{Li2018} proposed a framework that relies on a blockchain and a cryptocurrency, called CreditCoin, in order to motivate users to share information. However, the network latency to produce information and to notify the RSI is not evaluated.

Sharma \textit{et al.}~\cite{Sharma2017} defined a blockchain-based architecture that relies on a vehicle manufacturer or a transit department to issue and to revoke permissions for all vehicles. Vehicles act as regular nodes that produce information that is stored in the blockchain, and special miner nodes that are managed by the manufacturer/road transport authority are responsible for handling all requests/responses from regular nodes. The miner is the node responsible for creating new blocks containing the vehicles transactions and updating the peers.

Dorri \textit{et al.}~\cite{Dorri2017} proposed a blockchain-based framework to address security and privacy of smart vehicles. In their paper, they discuss multiple use cases including remote software updates and flexible automotive insurance schemes. However, their article does not provide a detailed technical discussion of their framework and neither do they propose solutions for addressing the issues that we focus on in this paper. 

Lunardi \textit{et al.}~\cite{lunardi2018} proposed a lightweight blockchain that is responsible for keeping the data produced in a smart home scenario. Their paper considers that each device in a smart home will produce information, and this data will be stored as a transaction in the device's block, thus each device has only one block attached to the blockchain. In their architecture, the blockchain is maintained in the gateways (which are devices with more resources than sensor/actuator devices) and each block can grow without any limit.

Table~\ref{tab:related} provides a comparative summary of the key aspects of relevant related works. As previously described, IOTA~\cite{iota} is a blockchain proposed to perform Machine-to-Machine (M2M) payments using IoT devices, however it uses Proof-of-Work (PoW) consensus, which requires a considerable time to append new information. Consequently, hardware requirement for a full node in IOTA is high and not compatible with IoT devices (\textit{e.g.}, a PC with at least 2GB of memory is recommended). Focusing in limited hardware, Dorri~\textit{et al.}~\cite{Dorri2017,dorri2017towards} and Lunardi~\textit{et al.}~\cite{lunardi2018} proposed blockchain-based solutions that achieve low latency for adding information to the blockchain, however, both proposals do not take into account the dynamism and privacy concerns that are important in smart city scenarios. While Li~\textit{et al.}~\cite{Li2018} and Sharma~\textit{et al.}~\cite{Sharma2017} presented solutions using blockchain for smart cities, their approach incurs long delays for adding and retrieving information to and from the blockchain.

\section{Blockchain Architecture}\label{sec:background}

\begin{figure*}[h!]
 \begin{center}
\includegraphics[width=0.85\textwidth]{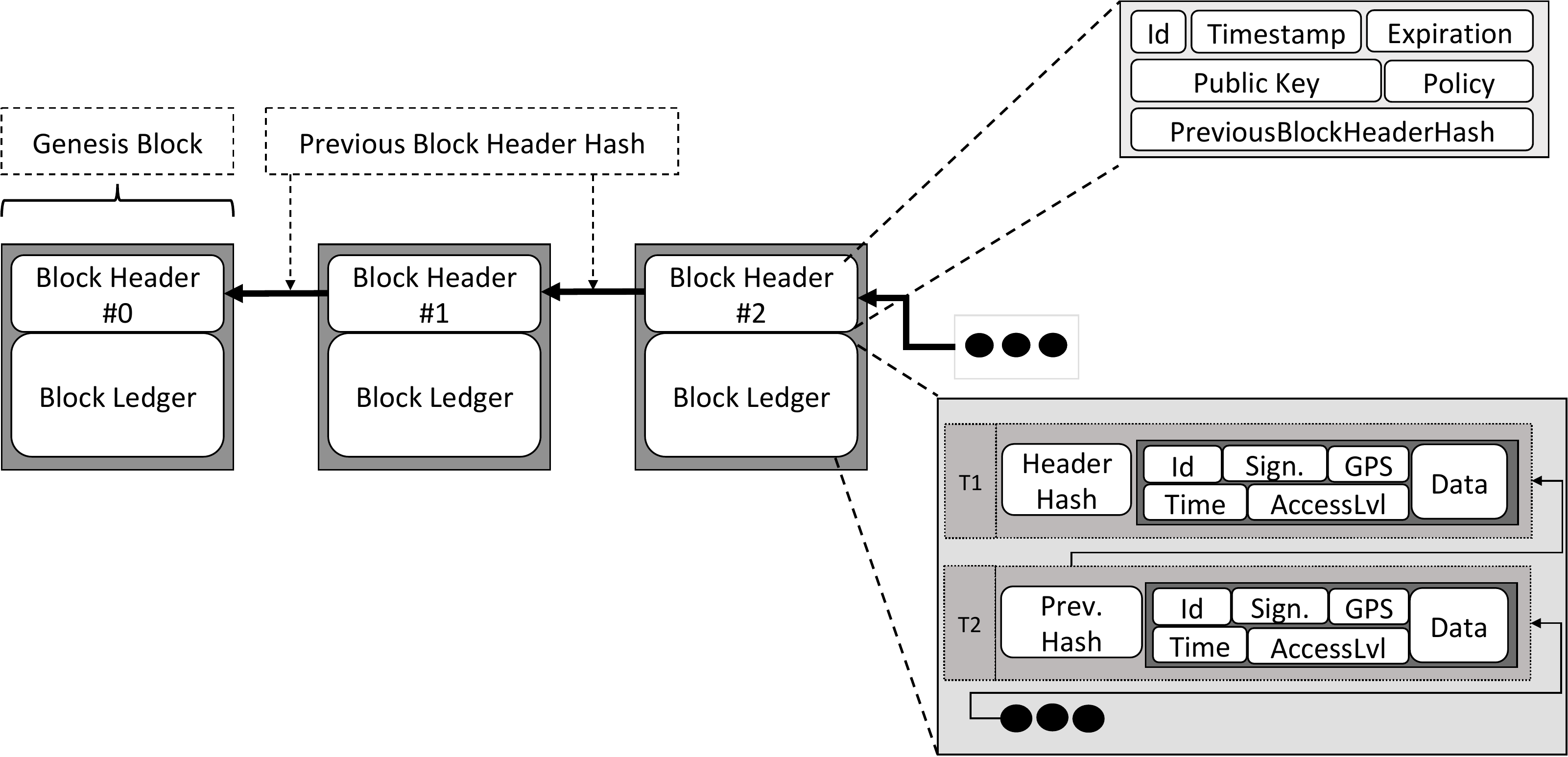}
 \end{center}
\caption{SpeedyChain data structure definition.}
\label{fig:blockchainFull}
\end{figure*}

In this section, we discuss the fundamental concepts of blockchain architecture that underpins our framework. The most prevalent blockchain implementations (\textit{e.g.}, Bitcoin and Ethereum) are not  suitable for SpeedyChain as they demand a lot of resources (computation and bandwidth), while in a smart city scenario some of the entities (\textit{e.g.}: traffic lights and parking spots) are resource constrained.  Additionally, some applications, such as vehicular networks, require low latency to information exchange. Furthermore, conventional blockchains require from a few seconds up to minutes to insert new information into their distributed ledger (depending on the used consensus algorithm). In a previous work, Lunardi \textit{et al.}~\cite{lunardi2018} presented a lightweight permissioned blockchain that creates blocks on demand. Therefore, each device will produce information and append data blocks to its own block. Consequently, devices will not have to wait for other devices to append its information into blockchain. Also, gateways are able to maintain only the Block Header - which contains the most important information about devices (especially their public keys).  It relies on signing the generated information and storing it in transactions that are attached to a block header. Each block contains two main parts as shown in Figure~\ref{fig:blockchainFull}. Here, we modify that approach to fit a smart city scenarios as follows: 

\begin{itemize}
    \item \textbf{Block Header}: the header is the only part of the block that is hashed and used in the header of the next block. This operation creates the link between the blocks in the blockchain. Thus, it should contain the necessary information to identify and verify that the transactions were produced only by the device that is the public key owner. We introduce an \textit{expiration} field that defines for how long (deadline) a block can accept new transactions. Therefore, this change prevents a block from growing in size indefinitely.  In the solution proposed by \cite{lunardi2018}, a block could accept new transactions at any time.
    
    \item \textbf{Block Ledger}: the payload contains the transactions that are chained to the block header. Unlike conventional blockchains where all transactions of the same node are chained together and might be in different blocks, SpeedyChain chains all valid transactions of the same node to the same block header, as shown in Figure~\ref{fig:blockchainFull}. As the link between the transactions is based on the block header, the transaction can be stored elsewhere, to avoid consuming space on the RSI, while making device-specific queries faster. Moreover, the inclusion of new information in each block could be performed in parallel. 
    
    \item \textbf{Sensor Data Support}: in order to support the inclusion of smart city data, the structure of adopted blockchain transactions was adapted to support data produced by the sensors of the vehicles (which could be speed, temperature, etc.). Similarly, we added fields to insert a \textit{geotag} (GPS position where the data were produced), an access level definition to describe the type of information stored in the transaction and the information about which user has permission to access the information.
\end{itemize}

Mining a block into the blockchain involves running a consensus algorithm that is a hard-to-solve easy-to-verify puzzle to protect the blockchain against malicious miners attempting to flood the blockchain with blocks. However, conventional consensus algorithms, \textit{e.g.}, PoW and Proof-of-Stake (PoS), incur significant (processing and memory) overheads, which are beyond the capabilities of IoT devices. To address this challenge, Lunardi \textit{et al.} \cite{lunardi2018} designated gateway nodes that are responsible for validating transactions and maintaining the blockchain. Besides gateway nodes, device nodes are responsible for gathering information from the environment, creating transactions with this information and sending them to the gateways. SpeedyChain works in a permissioned way using. It uses the structure as shown in Figure~\ref{fig:blockchainFull}, where, each block is created only after the RSI receives a witness (an entity that will ensure that the block was created by a genuine vehicle) validation, which means that it is guaranteed that the vehicle physically exists in the claimed area.

\begin{figure}[h!]
 \begin{center}
\includegraphics[width=0.48\textwidth]{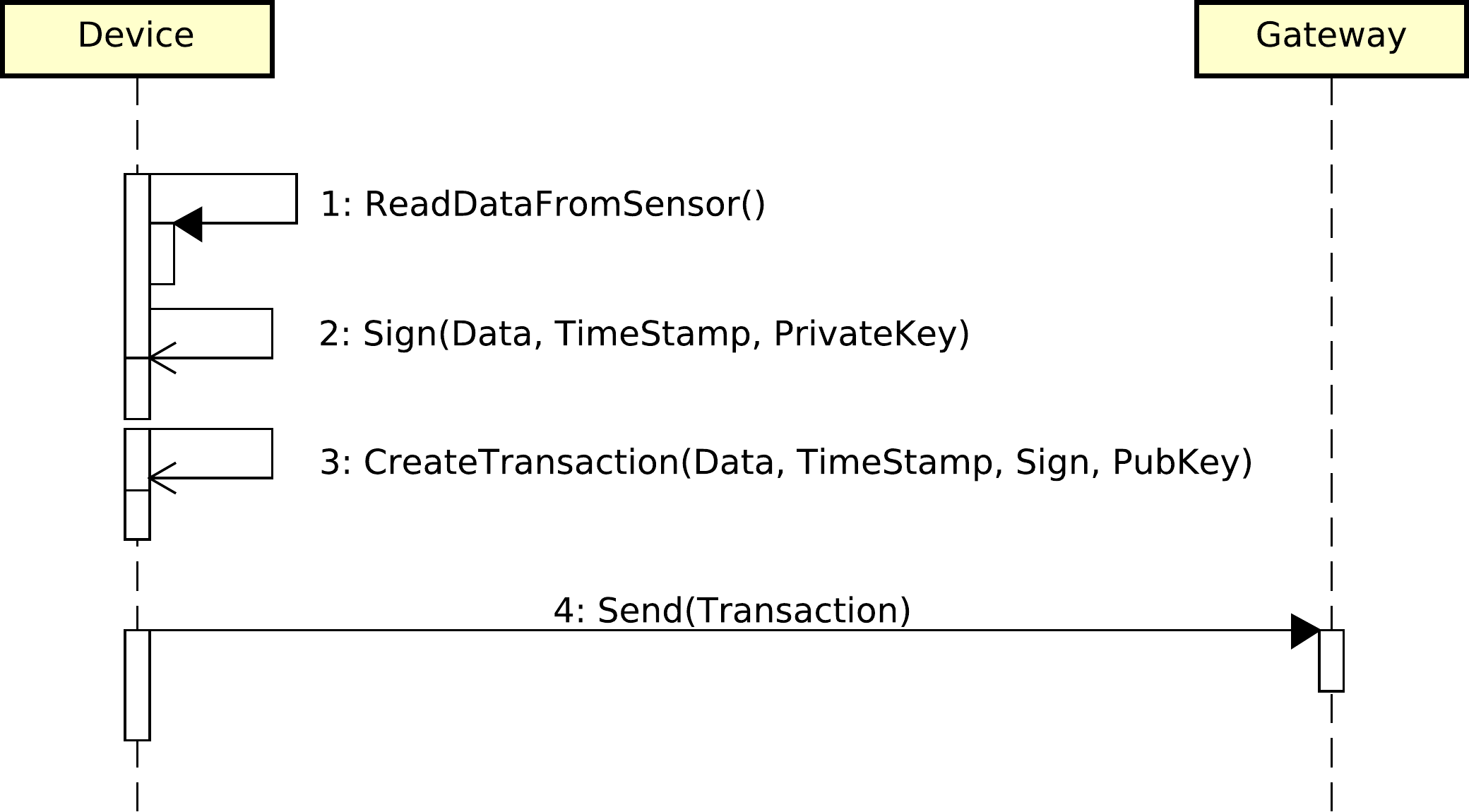}
 \end{center}
\caption{Device data generation.}
\label{fig:generateData}
\end{figure}

Figure~\ref{fig:generateData} presents the device data generation process, based on~\cite{lunardi2018}, and adapted for the smart city scenario. Each device is responsible for retrieving data from the environment (Step 1), signing the generated data (Step 2), creating a transaction (Step 3) and then sending this transaction to the gateway (Step 4). The gateway verifies whether there is a block in the blockchain associated with the device public key. Otherwise, the gateway will create a new block for the given device and update the peer gateways. It is important to clarify that each device corresponds to one block in the blockchain, to which all of its transactions are appended in its block ledger. Once the gateway has a block associated with the device, it must validate the transaction signature (to ensure that the information was created by the device), and if the signature is valid, the transaction is appended to the block (inside the block ledger) and sent to the peers.

\section{SpeedyChain Framework}\label{sec:architecture}

This section outlines the architecture of the proposed SpeedyChain framework. As shown in  Figure~\ref{fig:smartcity}, the involved entities can be vehicles, RSIs, and  SPs such as police stations, insurance companies, traffic management companies and digital maps providers like TomTom~\cite{tomtom2010}. Vehicles may use V2V or V2I communication~\cite{priv-ITS1}.  Furthermore, vehicles may generate data for one of two purposes: 

\begin{itemize}
    \item control data: data that are generated by each vehicle that can be used by other vehicles or traffic management and city planners to mitigate congestion in the city, \textit{e.g.}, the traffic report;
    
    \item service data: data that are sent to a particular node, \textit{e.g.}, an SP or vehicle owner. These data can be used to offer personalized services to the vehicle owner, \textit{e.g.}, flexible insurance, or can be used by the vehicle owner to monitor particular aspects of the vehicle. The user is responsible to decide which information will be available to each SP.
\end{itemize}

In the proposed architecture, the blockchain is managed by the RSIs and SPs as they have higher resources and less dynamism compared to vehicles. The vehicle itself is responsible for gathering data from its sensors, signing them and generating a new transaction, which is sent to the nearest RSI. This transaction should be validated by the RSI, as the RSI has access to the vehicle public key stored in the block header on the blockchain. A valid transaction is immediately appended to the current block of that vehicle or a new block is created if this is a new vehicle and broadcast to the peers as described in Section~\ref{subsec:operation}.  To ensure  vehicle privacy, the vehicle public key is changed at specific periods of time known as KUI. As vehicles have limited resources, they only need to maintain a Merkle tree of the blocks instead of keeping the whole blockchain (see Section~\ref{subsec:KUI}). 

\subsection{Initialization}\label{subsec:initialization}

The proposed architecture works in a permissioned mode, where only the SP and RSI have rights to manage the blockchain. While the vehicles are critical entities that generate data, and the RSI plays a crucial role in managing the blockchain, we propose to employ specific controls on how these entities can join and operate the blockchain:

\begin{figure}[h!]
 \begin{center}
\includegraphics[width=0.48\textwidth]{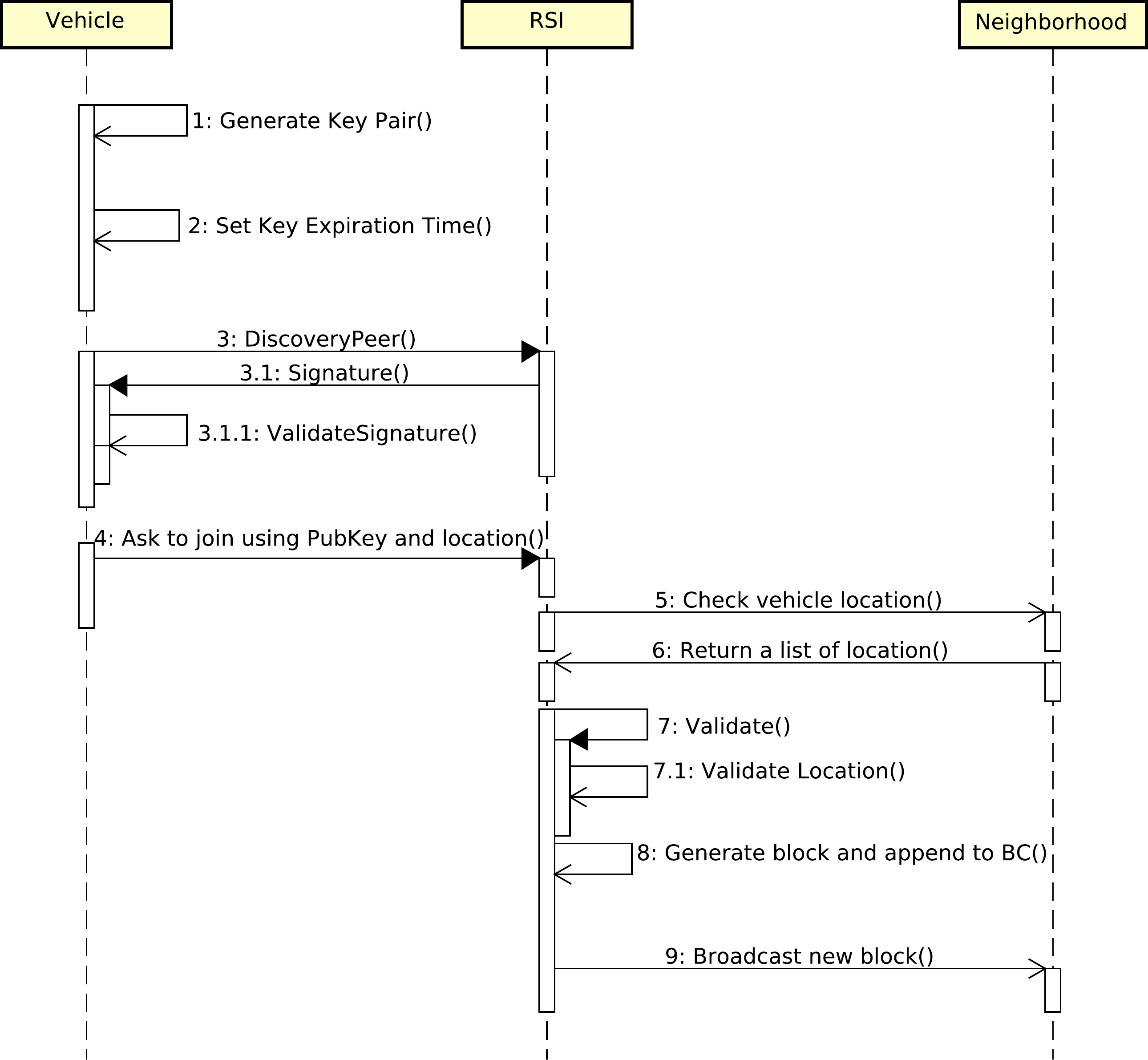} 
 \end{center}
\caption{Smart vehicle presenting to RSI.}
\label{fig:vehicleStart}
\end{figure}

\textit{Vehicles:}  for a vehicle to join the network it first must prove that it is a genuine vehicle, \textit{i.e.} it physically exists. This protects the network against Sybil attacks \cite{Dinger2006}, where a malicious vehicle pretends to be multiple vehicles and floods the networks with packets. To achieve this, we use the location-based trust model as proposed in~\cite{Brambilla2016}, which consists of gathering evidence from witnesses to support validation of the vehicle's existence\footnote{When the RSI does not receive any witness response, the vehicle block will be kept in a new block pool.}. In our framework, the witnesses are RSIs and the vehicle's neighbors at first connection to the network, which should obtain a location confirmation, from the unknown vehicle. Figure~\ref{fig:vehicleStart} outlines the main  steps required for a vehicle to  join the blockchain. Initially, the vehicle generates a new public/private pair (Step 1). The vehicle must set an \textit{expiration time} for its keys, which is the time period for each key is considered valid by other participants in the blockchain (Step 2).  The vehicle then populates a genesis transaction (first transaction in the block ledger) with its public key and location and sends it to the nearest RSI (Step 4). If there is no reachable RSI, the vehicle must wait till it is in the proximity of one. Note that this is a one time process for joining the blockchain. On receiving  the genesis transaction, the RSI must first verify that the vehicle is genuine. To do so, the RSI sends a request to other vehicles and RSIs near the vehicle's location (which is extracted from the genesis transaction), asking them whether such a vehicle exists in their vicinity (Step 5). This confirmation prevents the possibility of Sybil attacks. Once the RSI receives witness reports from peer vehicles (Step 6), it proceeds to validation (Step 7). Once the validation is successful, a new block is created (Step 8) and broadcast to all RSI peers (Step 9). Thus, the vehicle now is able to start appending new transactions to the block ledger in its block. 

\textit{RSI:} it is important that the vehicles or other entities in the smart city trust the RSIs. For this reason, each new car is registered to at least one RSI when these vehicles are joining to the network. After this initial process, the vehicle is able to move between cities. However, when it enters a new city, the vehicle may initially  have no trust in the RSIs. In our framework, each city has a unique block header created by the government (or other authorities). In the blockchain, each vehicle should keep only one active public key, where its data are stored. During the authentication phase, the RSI sends its signature (Step 3.1), which will be validated by the vehicle and when it is verified (Step 3.1.1), the vehicle asks to join the blockchain (Step 4), as shown in Figure~\ref{fig:vehicleStart}.

\subsection{Operation} \label{subsec:operation}

The information produced by vehicles is the main asset in the architecture. Thus, protecting the integrity of the data and the trust in the data generator, as well as the privacy of the data generator, are crucial in smart cities, as supported by the transaction definition in Section~\ref{sec:background}.

\begin{figure}[h!]
 \begin{center}
\includegraphics[width=0.48\textwidth]{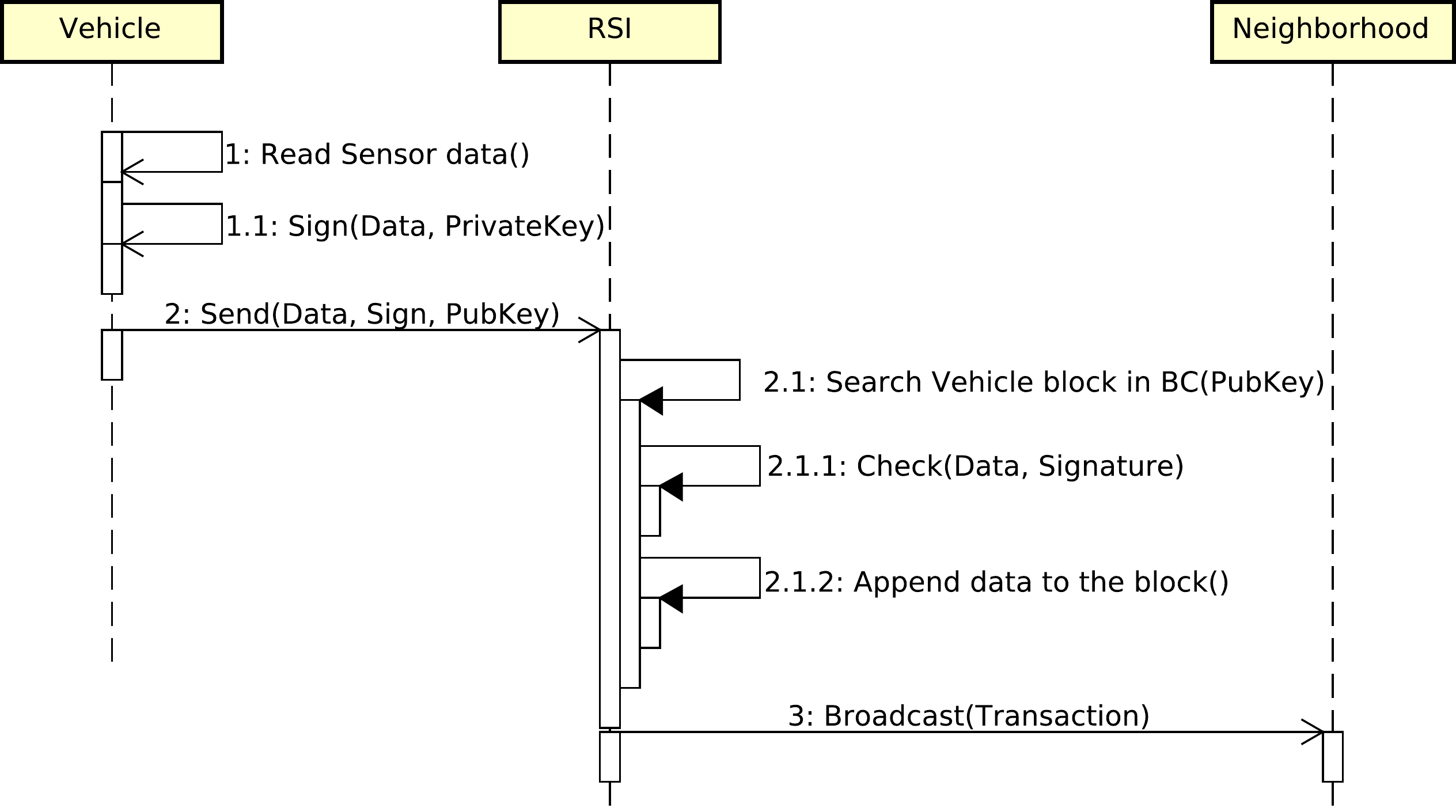}
 \end{center}
\caption{Sensor Data upload by the vehicle.}
\label{fig:vehicleOperation}
\end{figure} 

Figure~\ref{fig:vehicleOperation} depicts the typical process whereby a vehicle uploads sensor data. The vehicle reads data from its sensors (Step 1), signs these data transactions using its own private key (Step 1.1) and sends this transaction to the nearest RSI (Step 2). When there is no RSI nearby, the transaction is sent to the nearest vehicle and forwarded  until an RSI is found. The RSI then locates the vehicle block in the blockchain using its PK (Step 2.1). The RSI must then verify  the signature in the transaction (Step 2.1.1). In case of success, the transaction is appended to the block (Step 2.1.2) and broadcast to all peers (Step 3).

In addition to the regular operation, there are two other scenarios that vehicles could face when producing information: \textit{i}) unreachable RSI: in this case, the vehicle that is creating a transaction cannot be reached by other RSIs or vehicles. Thus, the vehicle will continue storing its data in transactions and will send those transactions to the RSI when it is connected again. During this time, its information in the blockchain will be outdated, until receiving the transactions; and \textit{ii}) Vehicle-to-Vehicle communication: in this case, the vehicle can only communicate with another vehicle. Thus, it will send the transaction to its neighbor vehicle to be forwarded to the RSI, as soon one of them reach an RSI. This effectively makes the nearby vehicle a transaction mule to the nearest RSI.


\subsection{Key Update Interval (KUI)} \label{subsec:KUI}

Each vehicle is authenticated using a Public Key (PK), and, this PK is available to any entity in the architecture through the blockchain. This usage is similar to the research presented in the Bitcoin blockchain~\cite{Nakamoto2008}, where each wallet is identified by its PK, and, is recommended that the users change this PK from time-to-time in order to improve its security. In our architecture, this key is regularly changed to prevent malicious nodes from identifying the vehicle. To address this challenge, in our architecture the vehicles can change their PK by storing a new block header in the blockchain. This implies that all participants must store an updated view of the blockchain.  However, in a smart city, some entities might have limited storage capacity, \textit{e.g.}, smart vehicles. To decrease the memory overhead on these entities, the low resource devices only store the root hash of a Merkle tree constructed using the PK of nodes in the network. Each vehicle can change its PK, which subsequently will change the Merkle tree. Thus, vehicles must periodically generate the newest Merkle tree. The RSIs are also responsible for rebuilding the Merkle-tree at the end of a periodic time interval known as the KUI. 

The RSI stores the PK in the new block headers mined in the blockchain during the KUI. At the end of the KUI, the RSI constructs a Merkle tree based on the block header. Then, it broadcasts the calculated value to every node in the network. By receiving the notification, vehicles update the root of the Merkle tree. Additionally, each node stores the necessary information to prove its membership in the Merkle tree so that it later can prove that its PK is a member of the Merkle tree and thus authenticate itself. 

\section{Evaluation and discussion}\label{sec:evaluations}

In this section we first provide a qualitative discussion on the security and privacy of our method (Section \ref{subsec:secuirty-privacy}) followed by performance evaluation results (Section \ref{subsec:performance-evaluation}). \par 

\subsection{Security and privacy analysis}\label{subsec:secuirty-privacy}

We assume that the adversary (or cooperative adversaries) can be any node in the SpeedyChain framework. Adversaries are able to sniff communications, discard transactions, create false transactions and blocks, change or delete data, analyze multiple transactions in an attempt to compromise a node, and sign fake transactions to legitimize colluding nodes. We assume that standard secure encryption methods are used between the involved entities, which cannot be compromised by adversaries.\par 

\textit{Security:} We define the following possible attack scenarios: \par 

\begin{itemize}
\item \textbf{Sybil:} in this attack, the malicious node creates multiple identities for itself to either flood the network with transactions or create false claims, \textit{e.g.}, false traffic jams.  To perform this attack, the attacker pretends to be multiple vehicles and uses different identities, \textit{i.e.}, public/private keys. The proposed architecture is resilient against this attack as the vehicles authenticate the generator of a transaction before processing the transaction. Thus, the attacker will require the public/private keys of multiple vehicles that exist in the network and cannot fake identities.  \par 

\item \textbf{Modifying the data/transactions: } in this attack, the attacker modifies the content of transactions or the data that are sent in a transaction. Recall from Section~\ref{sec:background} that each newly created transaction is signed (in particular the hash representing the transaction and its content) by the node creating the transaction before it is sent to the blockchain and stored on it. Altering the content of the transaction could be detected by any network peer by checking its signature, and thus the transaction is discarded. \par 

\item \textbf{Malicious RSI: }A compromised RSI will continue to receive data from vehicles. However, it is not able to change any information stored within the transaction, because the vehicles are responsible for signing the transactions, and their private key are secure.  \par
\end{itemize}

Traditional attacks such as Majority, Race, Selfish Mining, and Block-Withhold~\cite{Conti2018}, exploits the PoW consensus algorithm, and as SpeedyChain does not rely on the PoW, these attacks were not considered as a threat to the present research. In the same way, Transaction Malleability attack~\cite{Conti2018}, which exploits the mechanism on how the Bitcoin transactions are generated, particularly the transaction id field, does not affect SpeedyChain.

\textit{Privacy:}  privacy in our proposed architecture, as in the Bitcoin blockchain, is mainly inherited from the public key mechanism that identifies each participating entity. However, in our proposal, the usage of each PK for each entity is limited by its expiration time, which when reached forces the entity to regenerate a new key pair.

\subsection{Performance evaluation}\label{subsec:performance-evaluation}

In this section we present results in an emulated scenario to evaluate the performance of our approach. We evaluated the network using the CORE network emulator~\cite{CORE2008}, running in a VM in VirtualBox with 4 processors and \textit{8GB} of \textit{RAM} and the host Intel \textit{i7@2.8Ghz} and \textit{16GB} of \textit{RAM}. The emulated network consists of 15 nodes representing RSIs that are interconnected with other RSIs and keep the blockchain updated. Thus, an RSI will be responsible for receiving connection requests from the vehicles, validating them, creating new blocks, and broadcasting these blocks to the other RSIs (as shown in  Figure~\ref{fig:vehicleStart}).  Vehicles are responsible for establishing a connection to an RSI and, once connected, produce data and send that to the RSI (as shown in Figure~\ref{fig:vehicleOperation}).

To evaluate the scalability of the proposed method, we vary the number of transactions generated by the vehicles, and study the network performance with 10, 100, and 1,000 transactions in the network. We also vary the number of participating vehicles in the network and study the performance in three scenarios  with 50, 100, and 650 vehicles. Overall we run 9 different scenarios to study the performance of each metric, that allows evaluating the proposed framework performance to manage transactions and blocks in the smart city scenario and still keeping this time in a few milliseconds according to Table~\ref{tab:related} presented in Section~\ref{sec:relatedWork}.

\begin{figure}[h!]
 \begin{center}
\includegraphics[width=0.475\textwidth]{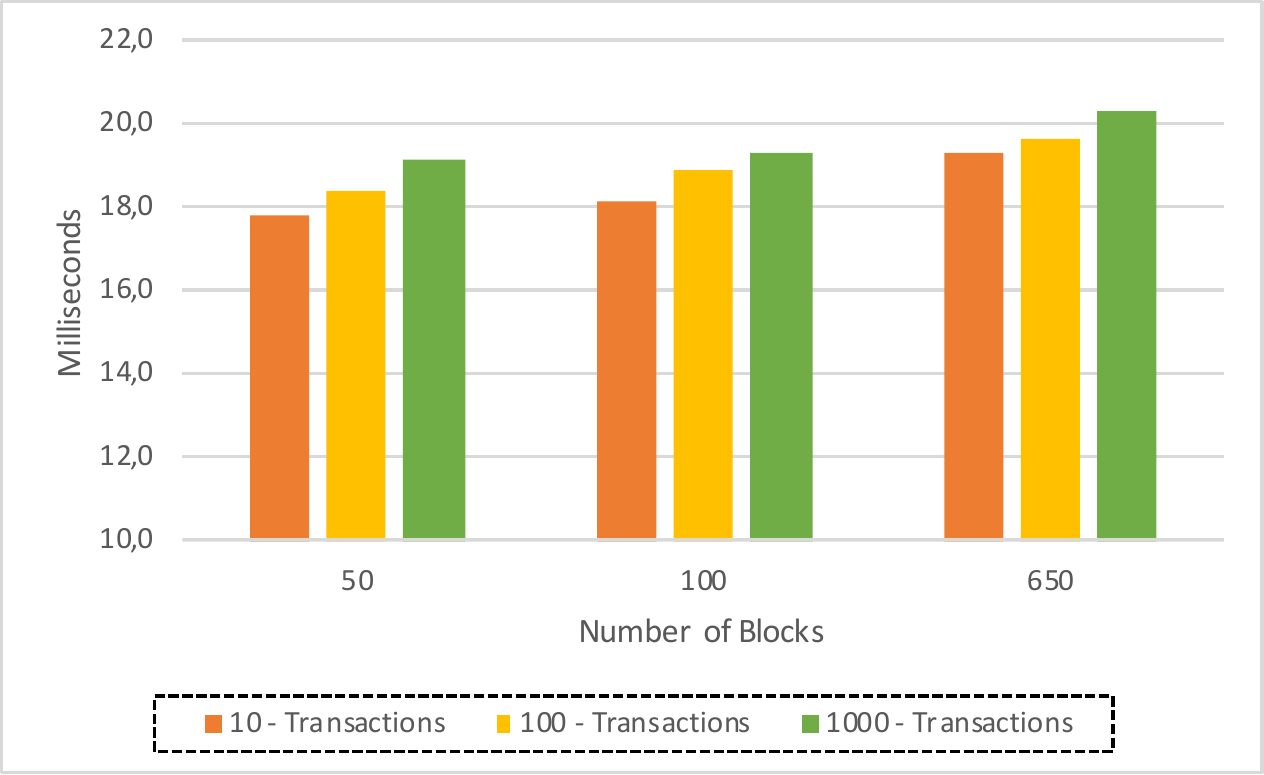}
 \end{center}
\caption{Required time in an RSI to add a new block to the blockchain.}
\label{fig:RSInewBlock}
\end{figure}

\begin{figure}[h!]
 \begin{center}
\includegraphics[width=0.475\textwidth]{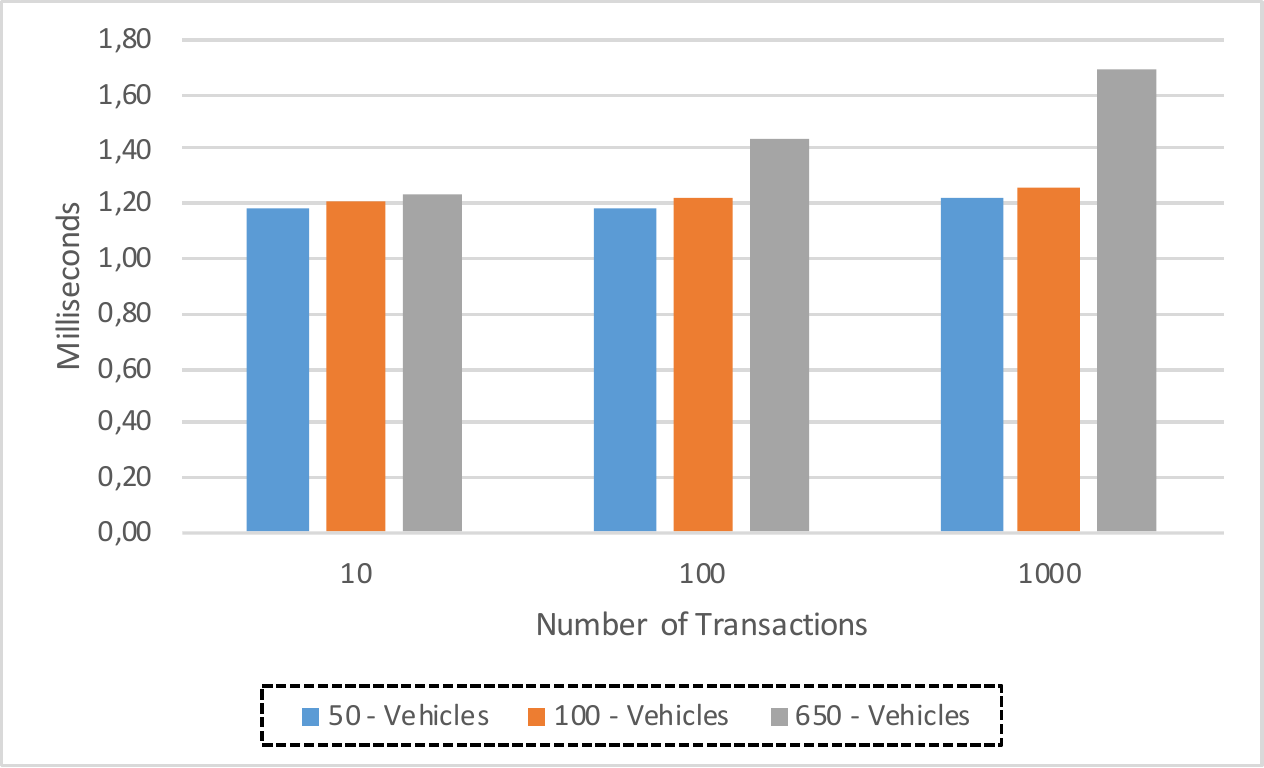}
 \end{center}
\caption{Required processing time to add new transaction to the blockchain.}
\label{fig:RSInewTrans}
\end{figure}

We first evaluated the processing time taken by an RSI to add a new block to the blockchain. The process of adding a new block includes: \textit{i}) receiving a connection request at the RSI, \textit{ii}) validating the vehicle request, \textit{iii}) identifying the need for a new block and creating it, and \textit{iv}) updating the RSI peers.  Figure~\ref{fig:RSInewBlock} summarizes the emulation results. As expected, the processing time for adding a new block to the blockchain increases as the number of blocks increases since there are more transactions and blocks to be validated by the RSI. It can also be seen that the number of transactions also directly affects the processing time (Figure \ref{fig:RSInewTrans}). The reason is that a higher number of transactions in the block ledger requires longer to validate. However,  processing time overhead for transactions is less significant than for new block creation. Note that the block creation operation will only be executed when a vehicle connects to an RSI the first time (or when it changes its key pair). 

The block is created only when a vehicle joins the network. However, once its block is added to the blockchain, it will be allowed to generate transactions. The next metric that we evaluated is the time taken by RSIs to process received transactions, \textit{i.e.}, verify data signature, check if the block is not expired, append the transaction and notify all RSIs regarding the update. Figure~\ref{fig:RSInewTrans} plots this metric. In this scenario, it was noticed that for a blockchain size of 50, the time increases 3.63\% when increasing the number of transactions from 10 to 100. In a blockchain with 100 blocks, the transaction creation time increases 4.62\%. However, for a blockchain size of 650 blocks, the time to validate and append a new transactions increases by 36.97\% from 10 to 1,000 transactions, which points to a linear time increase, as expected. Based on the collected samples, assuming a confidence level of 95\%, the interval varies from 0.0013 ms to 0.0126 ms, which lead us to a very small variance on the data.

\begin{figure}[h!]
 \begin{center}
\includegraphics[width=0.475\textwidth]{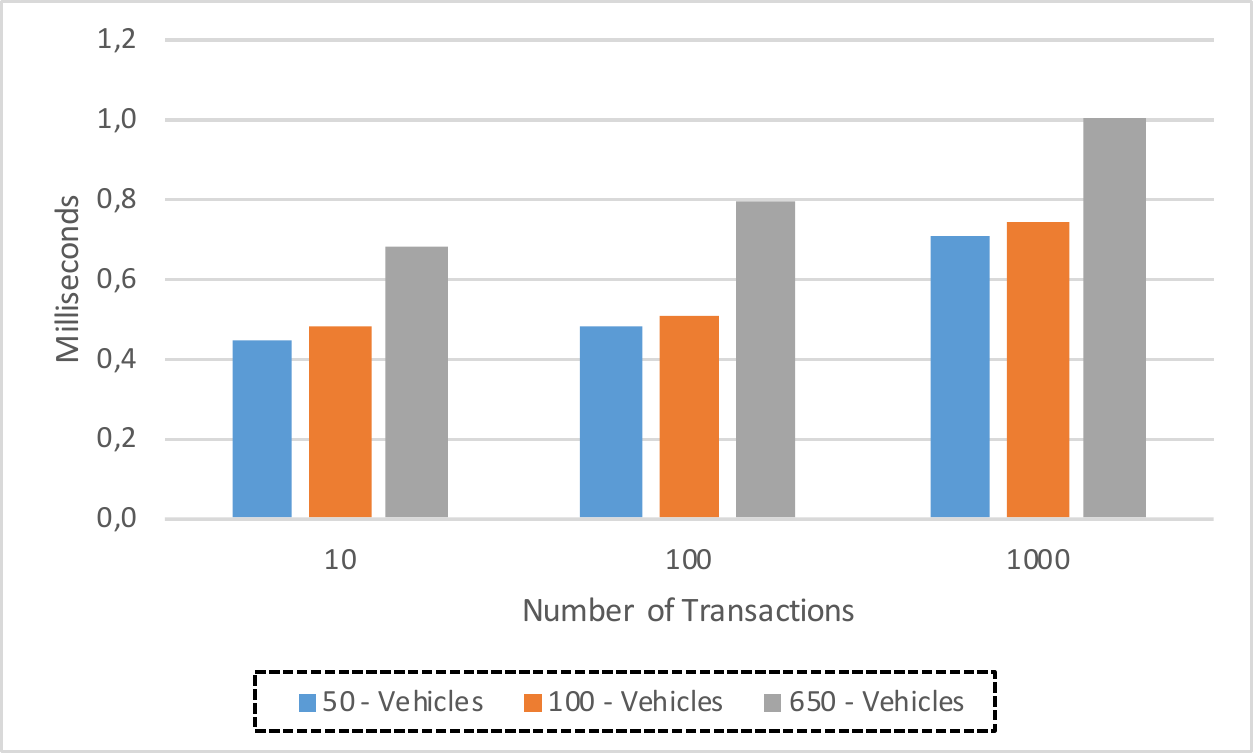}
 \end{center}
\caption{Time (ms) to update peer's blockchain with received transactions}
\label{fig:peerAddTrans}
\end{figure}

Each transaction is created by the vehicle and sent to the RSI. The receiving RSI  validates the transaction to ensure the integrity and trust on the data coming from the vehicle. Once all validation is performed, the RSI updates its local blockchain copy and  sends it to other RSIs. Each RSI that receives the new transaction validates the received data before appending this new transaction to local copy of the blockchain. The time required to execute this update operation increases as more transactions are stored in each block, and this behavior is shown in the tests presented in Figure~\ref{fig:peerAddTrans}. Considering a blockchain with 650 blocks and changing the number of transactions from 10 to 1,000, the time to validate and update the blockchain  increases by 103.08\%, which is justified by having 100 times more transactions to be processed, which lead us to a linear time grow. Considering a confidence level of 95\% for the samples, the error probability is at 0.003 ms to 10 transactions and 0.002 ms to 1,000 transactions.

\begin{figure}[h!]
 \begin{center}
\includegraphics[width=0.475\textwidth]{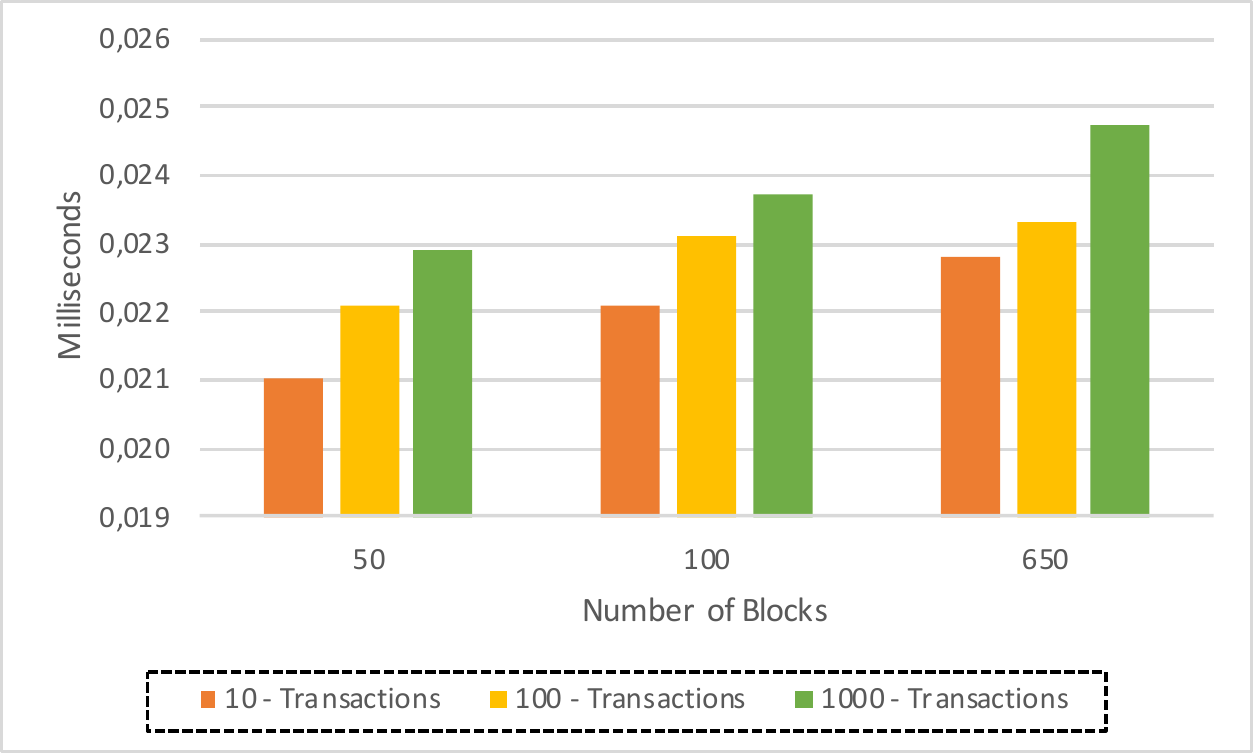}
 \end{center}
\caption{Time (ms) to update peer's blockchain with received blocks}
\label{fig:peerAddBlock}
\end{figure}

The time that a single peer takes to validate and to update its blockchain with a new block created for another RSI is shown in  Figure~\ref{fig:peerAddBlock}. This value varies between 0.021 ms and 0.025 ms for a blockchain with 50 blocks and 10 transactions to 650 blocks and 1,000 transactions, respectively.

Considering the evaluated blockchain sizes in terms of transactions and amount of blocks, even considering the scenario of a blockchain with 650 blocks sending 1,000 transactions, the values are 20.33 ms to validate and to create a new block and 1.69 ms to create and to validate a new transaction. The validation time for this scenario represents 7.7\% of the operation total time. It represents a good improvement in terms of time to add a new transaction in comparison with the Bitcoin blockchain.

\begin{table}[ht]
\centering
\caption{Time taken by vehicles to calculate the Merkle tree root.}
\label{tab:vehicleMerkle}
\begin{tabular}{|l|l|}
\hline
{\# of Transactions} & { \begin{tabular}[c]{@{}l@{}}Time to calculate \\ the Merkle tree\end{tabular}} \\ \hline
10                                        & 0.162 ms                                                                                        \\ \hline
100                                       & 0.857 ms                                                                                        \\ \hline
1000                                      & 7.995 ms                                                                                        \\ \hline
\end{tabular}
\end{table}

The time taken to calculate the Merkle tree in a vehicle was measured and is presented in Table~\ref{tab:vehicleMerkle}. As this operation is performed on the transactions, we consider three values for the number of transactions in the block ledger that are 10, 100 and 1,000, and evaluated the time to generate the Merkle tree for each set. The time for a block with 10 transactions was measured as 0.126 milliseconds, while for 1,000 transactions this time increases to 7.99 milliseconds. As expected, the time to calculate the Merkle tree increases as there are more transactions within a block. These results present a sublinear time increase, and, taking these amount of data, allows to act as input in order to estimate the KUI used to define the block expiration time.

\section{Conclusion}
\label{sec:conclusion}
Future smart vehicles will be increasingly connected and share information with smart city facilities, \textit{e.g.}, road side infrastructures, other vehicles, and service providers, to provide personalized services and to assist the driver for avoiding congestion, or to alert drivers about a dangerous situation. However, as the vehicles will exchange privacy-critical data, the privacy of the vehicle owner must be taken into account while designing a communication model for vehicles. 

This paper proposed SpeedyChain, a new blockchain-based framework to provide a private and secure communication model for vehicles. It also ensures data integrity  using hashes in transactions as well as the communication auditability using transaction records in the blockchain. Using the proposed framework, the smart city participants, including vehicles and roadside infrastructures, can trust that the data they received are generated by a trusted node. We built a two-way trust, meaning that the vehicle can trust other vehicles or RSIs, and  RSIs can also trust other vehicles. To ensure privacy, a vehicle can change its key at predefined time-intervals. Blockchain is used to ensure the identity of  the generator of any incoming transaction. The storage of the blockchain might not be applicable for low resource participants, \textit{e.g.}, smart vehicles, and should be discussed separately. Thus, in our approach vehicles only maintain the root hash of a Merkle tree of valid PKs in the network provided by a trust agent known as Merkle tree updater. 

We provided qualitative analysis on security and privacy of our approach. Simulation results using the CORE simulator demonstrate that the proposed method is capable of maintaining the blockchain on the RSI and the latency to create blocks and transactions, as well notify the RSIs about the change, is low. We also identified that as the number of transactions generated or blocks created in the blockchain increases, the time for performing these operations increases too, however, a further research will be conducted in order to evaluate the SpeedyChain scalability in a larger scenario.

An interesting direction for future work is to define different levels of access control for the data that are produced by a smart vehicle, allowing it to define the access level. Similarly, future work  should evaluate a mechanism to allow the transactions to be stored externally to the blockchain while maintaining its data sequence, integrity and non-repudiation attributes.

\section*{Acknowledgment}
This paper was achieved in cooperation with HP Brasil using incentives of Brazilian Informatics Law (Law n 8.2.48 of 1991). We also thank CAPES for the financial support. 

\bibliographystyle{IEEEtran}
\bibliography{Arxiv}

\end{document}